\documentclass[aps,prc,twocolumn, 10pt,groupedaddress]{revtex4-2}
\usepackage{xspace}

\newcommand {\sqrtsNN}  {\ensuremath{\sqrt{s_{_{\text{NN}}}}}\xspace}

\newcommand{\kt}{\ensuremath{k_{\text{T}}}\xspace}
\newcommand{\pt}{\ensuremath{p_{\text{T}}}\xspace}
\newcommand{\PYTHIA}{\textsc{pythia}\xspace}
\newcommand{\HERWIG}{\textsc{herwig}\xspace}
\newcommand{\PbPb}{\ensuremath{\text{PbPb}}\xspace}
\newcommand{\pp}{\ensuremath{\text{pp}}\xspace}
\newcommand{\DRaxis}{\ensuremath{\Delta\text{j}}\xspace}
\newcommand{\GeV}{\ensuremath{\text{GeV}}\xspace}
\newcommand{\TeV}{\ensuremath{\text{TeV}}\xspace}
\newcommand{\Raa}{\ensuremath{R_{\text{AA}}}\xspace}
            
\usepackage{graphicx}
\usepackage{booktabs}
\usepackage{makecell}
\usepackage[colorlinks=true,allcolors=blue]{hyperref}

\begin{document}

\title{Constraining color-charge effects of partonic energy loss with jet axis-based inclusive jet substructure measurement}

\author{Raghunath Pradhan}
\email[]{raghunath.pradhan@cern.ch}
\author{Olga Evdokimov}
\email[]{evdolga@uic.edu}
\affiliation{Department of Physics, 845 West Taylor Street, Chicago, \\ University of Illinois Chicago, 60607, USA}

\begin{abstract}
This study investigates the color-charge dependence of parton energy loss in the quark-gluon plasma (QGP) medium and the associated relative modifications of quark and gluon jet fractions compared to vacuum, using jet axis decorrelation observables. Recent CMS jet axis decorrelation measurements in \PbPb collisions at 5.02 TeV are interpreted using Pythia simulations with varied quark/gluon jet compositions and emulated color-charge dependent energy loss. A template-fit procedure is employed to estimate the limits on gluon jet fractions in the published CMS data and average shift in jet momentum due to quenching for quark- and gluon-initiated jets traversing the QGP. The extracted gluon jet fractions and the estimated quark and gluon energy losses based on this study of jet axis decorrelations are found to be consistent with other model calculations based on inclusive observables. This work illustrates the use of jet substructure measurements for providing constraints on the color-charge dependence of parton energy loss and offers valuable insights for jet quenching models.
\end{abstract}

\maketitle

\section{Introduction}
Ultra-relativistic heavy-ion collision experiments aim to recreate a deconfined state of quarks and gluons, known as the quark-gluon plasma (QGP), under laboratory conditions to investigate the properties of this unique form of nuclear matter. Jets, collimated streams of particles produced in high-energy collisions, serve as versatile tools for probing the QGP~\cite{Apolinario:2022vzg, Cunqueiro:2021wls}. These jets are initiated by energetic quarks and gluons (partons) produced in high-momentum-transfer processes during the initial stages of the collision. The partons subsequently fragment into lower-energy partons, which then hadronize into final-state particles. As the outgoing partons traverse the QGP, they lose energy through interactions with the medium. These interactions result in medium-induced modifications to various experimental observables, collectively referred to as ``jet quenching''. Experimental evidence for jet quenching has been observed using multiple probes at both the BNL RHIC~\cite{PHENIX:2001hpc, STAR:2002ggv} and the CERN LHC~\cite{CMS:2021vui, CMS:2021nhn, ATLAS:2023hso, ALICE:2023waz}. Such probes include the suppression of high transverse momentum (\pt) particle yields, modifications to jet fragmentation functions, and changes in jet shapes. Some recent reviews on this subject can be found in Refs.~\cite{CMS:2024krd, ALICE:2022wpn, Apolinario:2022vzg, Cunqueiro:2021wls}.

In parallel with experimental studies of quenching phenomena, significant theoretical advances have been made over the past two decades in characterizing partonic energy loss in the QGP medium. The mechanisms understood as the primary contributors to the energy loss are medium-induced gluon radiation and collisional energy loss. Despite substantial progress on both theoretical and experimental fronts, the details of parton-medium interactions and the precise mechanisms governing parton energy loss remain unclear. Emerging techniques, such as jet substructure studies and the tagging (or selection) of quark and gluon jets, offer new insights into energy loss mechanisms~\cite{Kogler:2018hem, Vertesi:2024tdv}. Due to differences in Casimir color factors ($\text{C}_{\text{F}} = 4/3$ for quarks and $\text{C}_{\text{A}} = 3$ for gluons), gluon-initiated jets are expected to interact more strongly with the QGP and lose more energy compared to quark-initiated jets. This enhanced energy loss reduces the fraction of gluon jets within the surviving jet population in heavy-ion collisions at a given jet \pt compared to the vacuum case without a medium. Recent measurements suggest that the relative fractions of quark and gluon jets are modified in heavy-ion collisions compared to elementary proton-proton collisions, reflecting the differential energy loss due to the distinct color charges of quarks and gluons~\cite{mediumqg, eg:partonloss, Zhang:2023oid, ALICE:2019ykw}.

The CMS experiment attempted to study gluon-like jet fractions by applying the template-fit method to the jet-charge observable for an inclusive jet sample from lead-lead (\PbPb) collisions at a center-of-mass energy per nucleon pair (\sqrtsNN) of 5.02~\TeV~\cite{CMS:2020plq}. This measurement found no significant evidence of a decrease in gluon jets or a corresponding increase in quark jets. The jet-charge observable currently lacks theoretical predictions for the expected in-medium modifications. The ATLAS experiment measured the jet nuclear modification factor (\Raa) and estimated the fractional energy loss ($S_{\text{loss}}$) for photon-tagged jets, and compared these observables with those for inclusive jets~\cite{ATLAS:2023iad}. Since photon-tagged jets are dominantly quark-initiated, this comparison provides sensitivity to the color-charge dependence of parton energy loss~\cite{ATLAS:2023iad}. It was reported that the \Raa of photon-tagged jets is significantly higher, and their estimated $S_{\text{loss}}$ significantly lower, than those of inclusive jets, providing evidence that quark jets lose less energy than gluon jets in the medium. However, a detailed description of color-charge dependent energy loss in the QGP medium, along with the relative modifications of quark and gluon jet fractions, remains elusive and warrants further experimental and theoretical investigation through additional observables.

Recently, the CMS experiment measured a new jet substructure observable, the jet axis decorrelation (\DRaxis), for a similar data sample (inclusive jets from 5.02~\TeV \PbPb collisions)~\cite{CMS:2024koh}. This observable is predicted to be sensitive to the parton-medium interactions and provide constraints for energy loss models~\cite{angelJetaxis}. The study found that \DRaxis is narrower than predictions from various unquenched models, such as \PYTHIA~\cite{pythia82} and \HERWIG~\cite{Sirunyan:2020pqv}. This narrowing could be attributed to modifications in the internal structure of jets caused by interactions with the QGP medium, selection biases favoring less-quenched jets, and/or results from the color-charge dependence of partonic energy loss. In the latter scenario, the larger average energy loss shifts gluon jets further down in energy compared to quark jets, resulting in a different partonic composition of quenched and unquenched jet samples.

In this paper, we attempt to interpret the recent CMS jet axis decorrelation measurements~\cite{CMS:2024koh} by comparing them to {\PYTHIA}8~\cite{pythia82} simulations with varying relative quark/gluon fractions and emulated color-charge dependent jet energy loss. A template-fit approach is employed to estimate the lower limit on gluon jet fractions within the inclusive jet sample. The quark- and gluon-initiated jet templates are derived from inclusive jet samples in 5.02 \TeV \pp collisions simulated with the {\PYTHIA}8 event generator. Simultaneously, the average jet energy loss for inclusive, quark, and gluon jets is estimated by minimizing the chi-square $(\chi^{2})$ between the \DRaxis distributions measured by CMS and those generated by {\PYTHIA}8 with an applied \pt shift imitating energy loss. Two scenarios are considered. First, in a simplistic approximation, all jets are assumed to lose the same fractional energy, with each jet’s \pt shifted down by the same fraction of transverse momenta. Additionally, we explore the color-charge dependent energy loss scenario, for which quark and gluon jets are allowed independent shifts to account for their differing energy losses.

The paper is organized as follows. Section~\ref{sec:ana} outlines the analysis framework used to measure jet axis decorrelation. Section~\ref{sec:tempfit} discusses the template fits, while Section~\ref{sec:shifpt} focuses on estimating the \pt shifts for inclusive, quark, and gluon jets to interpret the CMS \DRaxis distributions. Section~\ref{sec:result} presents the results, followed by a detailed discussion. Finally, Section~\ref{sec:sum} summarizes the main conclusions of this work.

\section{\label{sec:ana}Analysis method and framework}
The jet axis decorrelation, \DRaxis, is the angular difference between the axes determined by the anti-\kt energy-weighted (E-scheme) recombination and the winner-take-all (WTA) recombination schemes~\cite{angelJetaxis, CMS:2024koh}. It is defined as
\begin{equation}
\Delta \mathrm{j} = \sqrt{\bigl(\eta^{\mathrm{E\text{-}scheme}} - \eta^{\mathrm{WTA}}\bigr)^{2} 
              + \bigl(\phi^{\mathrm{E\text{-}scheme}} - \phi^{\mathrm{WTA}}\bigr)^{2}}
\label{eq:DRaxis}
\end{equation}

Where $\eta^{\text{E-scheme}}$ ($\eta^{\text{WTA}}$) and $\phi^{\text{E-scheme}}$ ($\phi^{\text{WTA}}$) denote the pseudorapidity and azimuthal angle of the E-scheme (WTA) axis, respectively, which characterize the jet's direction. In the E-scheme jet recombination~\cite{Cacciari:2011ma}, the jet axis is determined by the four-vector sum of its constituent particles, representing a momentum-averaged direction of energy flow that includes contributions from soft particles. This method clusters jets by iteratively combining particle pairs into pseudo-jets, with the direction of each new pseudo-jet defined by the vector sum of the four-momenta of its constituents. The final four-vector defines the E-scheme jet axis in $(\eta, \phi)$ coordinates. Conversely, the WTA scheme emphasizes the hardest prong at each clustering step, making it more sensitive to collinear radiation. To determine the WTA axis, the jet is reclustered using the WTA algorithm~\cite{wta1, wta2} with the same constituents, where the direction of each new pseudo-jet aligns with the highest-\pt particle or pseudo-jet. This typically results in the WTA axis aligning with the direction of the hardest constituent. The measured \DRaxis distributions are typically normalized to the total jet number in each \pt interval, constituting a shape-only measurement.

The {\PYTHIA} v8.226~\cite{pythia82} Monte Carlo event generator, configured with the CP5 tune~\cite{pythiaCP5tune} and NNPDF3.1 parton distribution functions at next-to-next-to-leading order~\cite{nnpdf31}, is used to simulate particle production in \pp collisions. Following the CMS data analysis, final-state jets are reconstructed using the anti-\kt algorithm~\cite{Cacciari:2008gp} with a distance parameter of $R = 0.4$, as implemented in the \textsc{FastJet} framework~\cite{Cacciari:2011ma}. To match the acceptance of the experimental work, simulated jets are selected within the kinematic range of $120 < \pt < 300$ \GeV and $|\eta| < 1.6$. 

In {\PYTHIA}, jets can be identified (or tagged) based on the type of parton originating the shower, allowing differentiation between quark and gluon jets. According to {\PYTHIA}8 simulations, the inclusive jet sample in the transverse momentum range of this study consists of comparable contributions from quark and gluon jets in the absence of medium effects, with the gluon-jet fraction being approximately 53\% at $\pt = 120$ \GeV and gradually decreasing to about 45\% at $\pt = 300$ \GeV. The predicted color-charge dependence of partonic energy loss in the presence of QGP is expected to result in a more significant migration of gluon-initiated jets towards lower final-state energy values than quark jets. Since gluon jet \DRaxis distributions are significantly broader than those of quark jets, differences in energy loss (and corresponding migration rates) could also contribute to the narrowing of the jet-axis decorrelations observed in \PbPb data by the CMS experiment~\cite{CMS:2024koh}.

\begin{figure*}[htbp]
\centering
\includegraphics[width=0.49\textwidth]{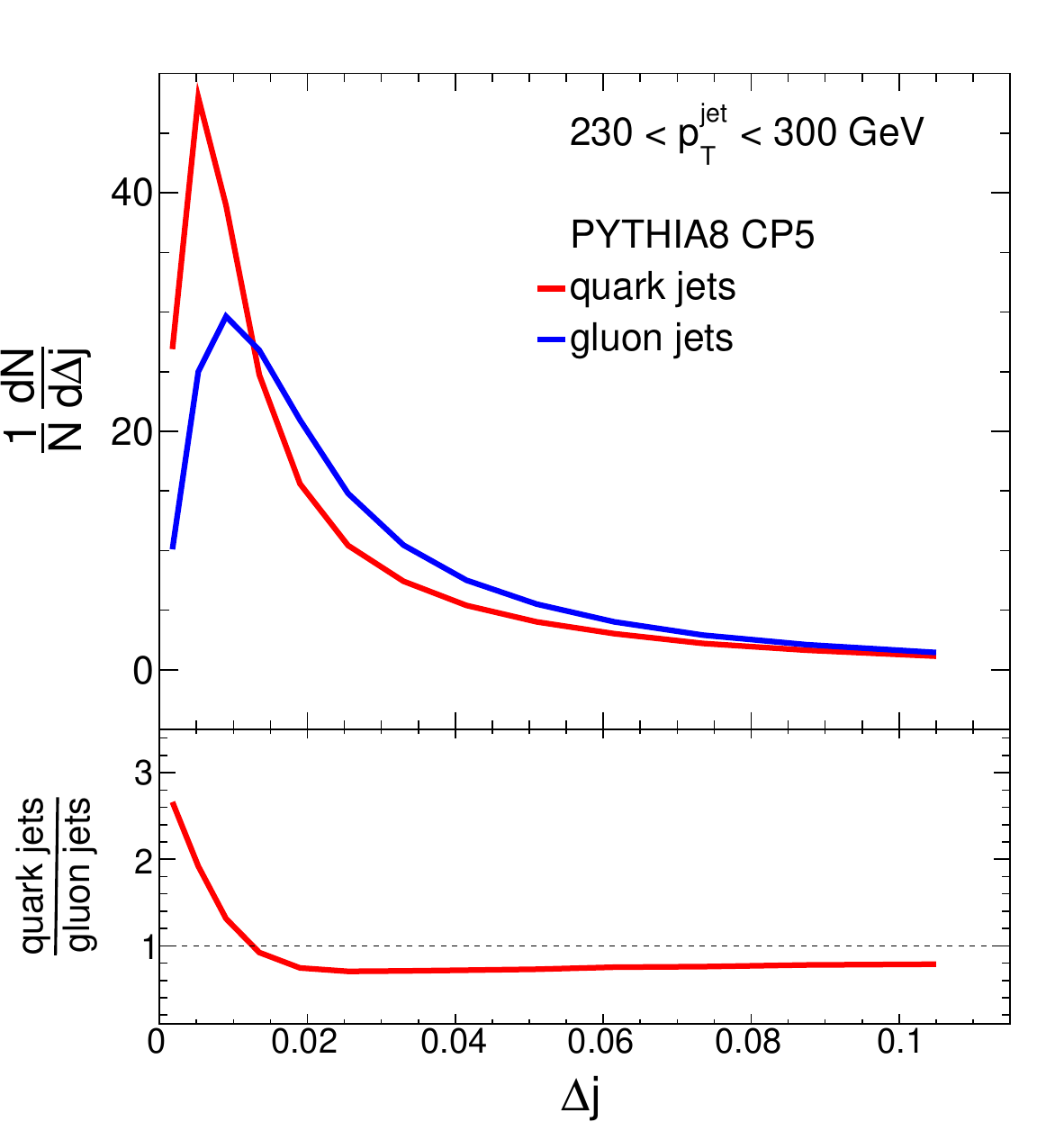}
\includegraphics[width=0.49\textwidth]{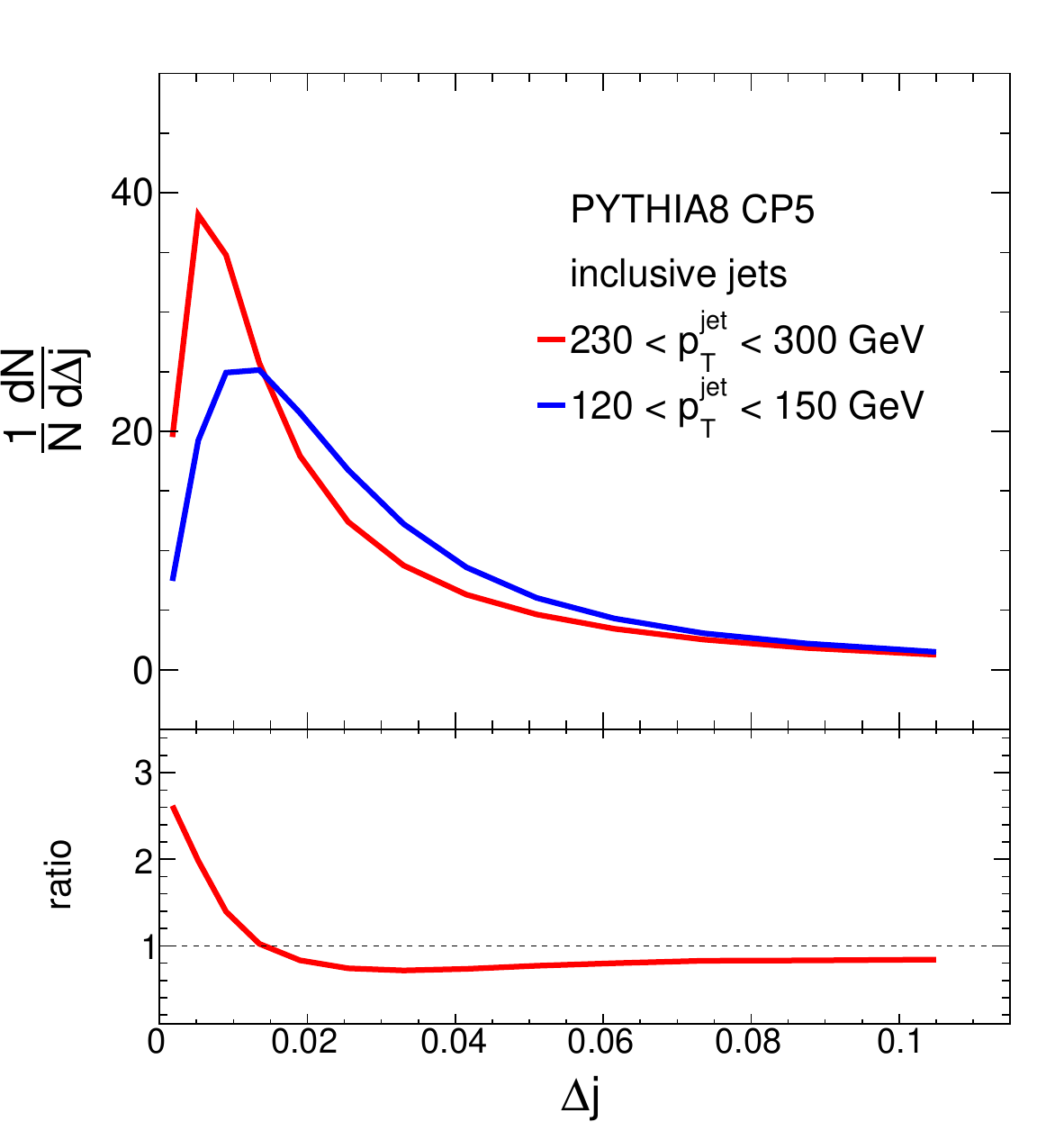}
\caption{\label{fig:flv_pt_shape_Dj}Left: $\Delta R_{\text{axis}}$ distributions for quark and gluon jets in {\PYTHIA}8, illustrating flavor dependence. Right: $\pt$ dependence of the distribution for inclusive jets. The bottom panel shows the ratio of the distributions.}
\end{figure*}

Figure~\ref{fig:flv_pt_shape_Dj} shows \DRaxis distributions from \PYTHIA simulations. The left panel compares quark- and gluon-initiated jets in the range $230 < \pt < 300$ \GeV, with quark jets exhibiting narrower \DRaxis distributions than gluon jets. The right panel illustrates the \pt dependence for inclusive jet distributions, comparing jets with $120 < \pt < 150$ \GeV and $230 < \pt < 300$ \GeV, indicating narrowing at higher-\pt. In both panels, the distributions exhibit stronger shape dependence at small \DRaxis, while at larger \DRaxis, there is little to no difference between quark and gluon jets or across different \pt intervals.

\subsection{\label{sec:tempfit}Template fits}
\begin{figure*}[htbp]
  \centering
    \includegraphics[width=0.495\textwidth]{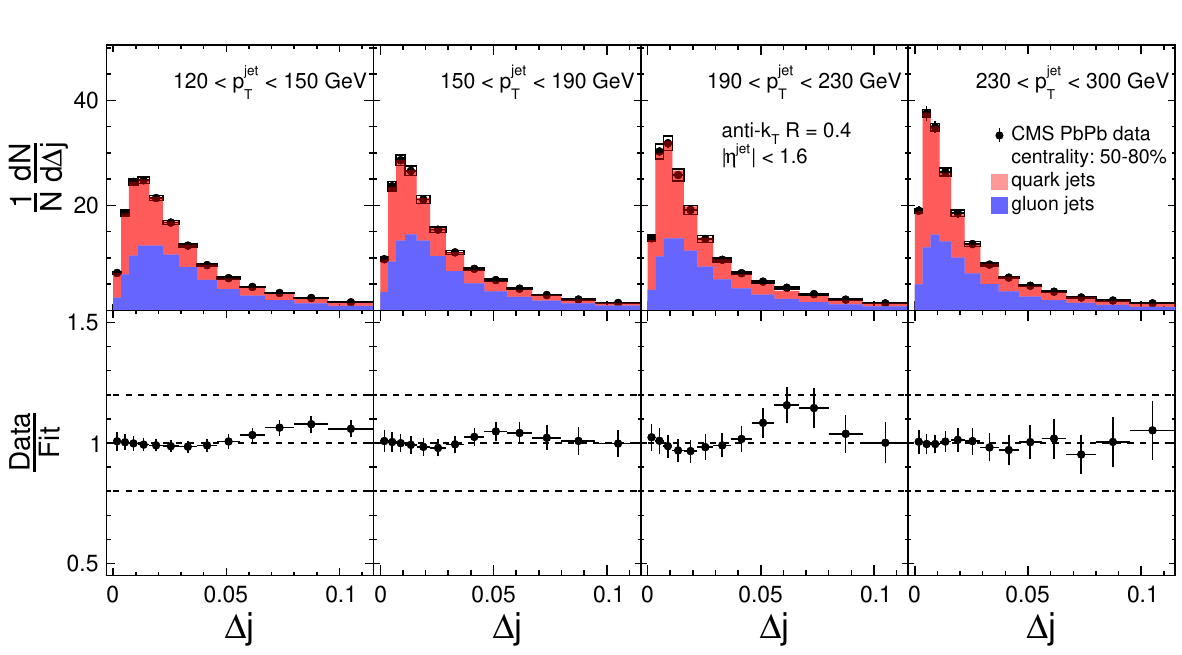}
    \includegraphics[width=0.495\textwidth]{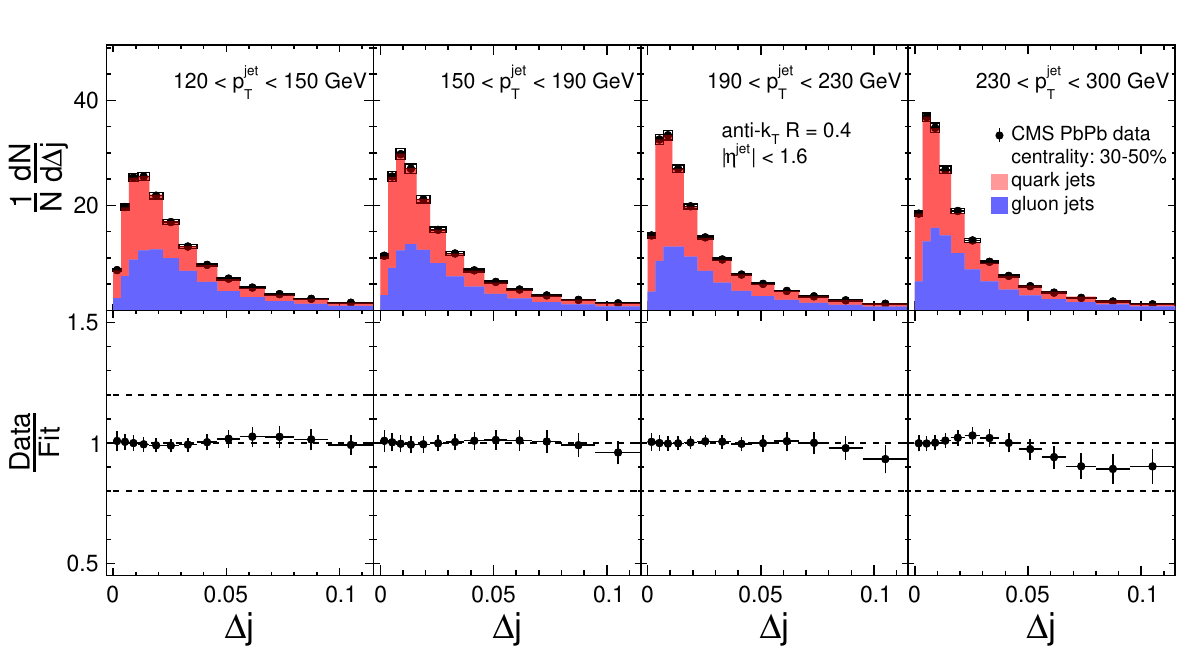}
    \includegraphics[width=0.495\textwidth]{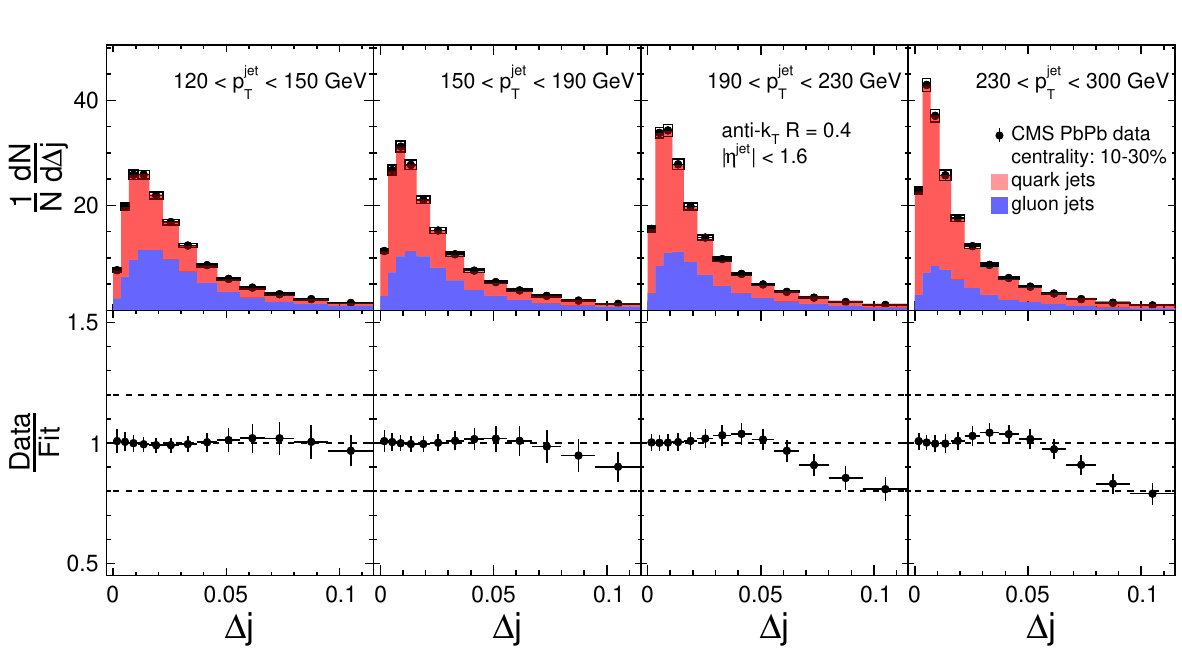}
    \includegraphics[width=0.495\textwidth]{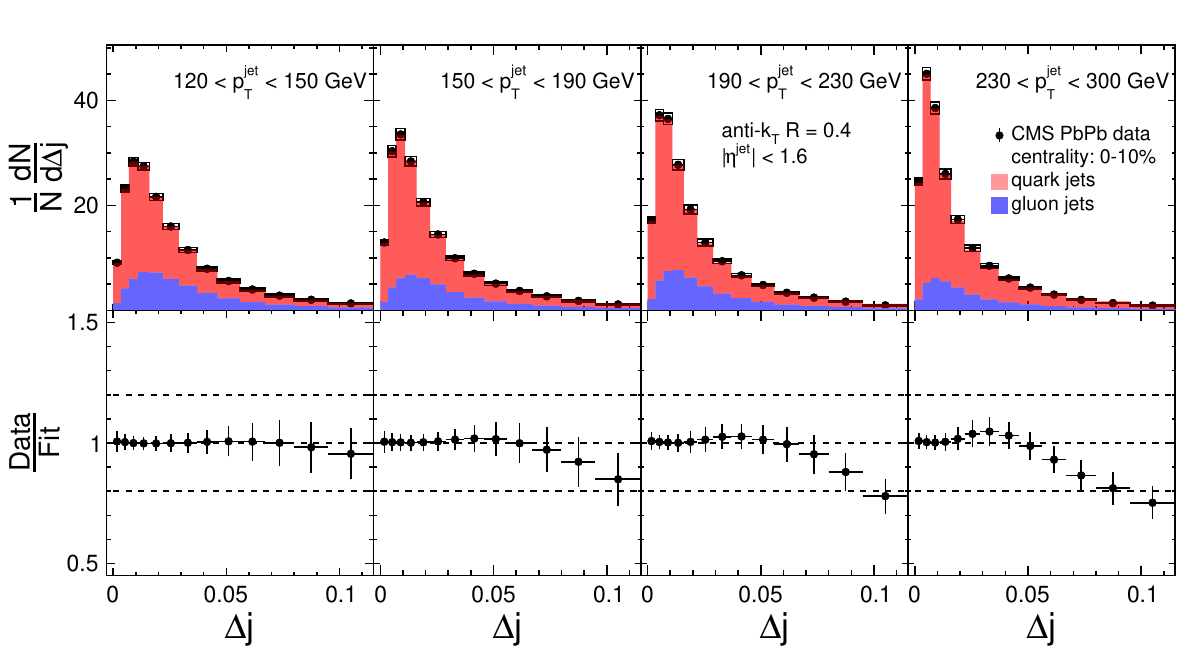}
\caption{\label{fig:tempFit_unfoldedCorr_pythia_cent3}
Template fits to the inclusive jet \DRaxis distributions (black circles) in \PbPb collisions for 50--80\% (top left), 30--50\% (top right), 10--30\% (bottom left), and 0--10\% (bottom right) centrality intervals, measured by CMS across different \pt intervals. Quark and gluon jet templates from \PYTHIA are shown as red and blue shaded areas, respectively. The lower panels show the data-to-fit ratios. Systematic uncertainties in CMS data are represented by rectangles (correlated) and shaded areas (uncorrelated), while vertical solid lines denote statistical uncertainties. In the lower panels, vertical lines represent total data uncertainties.}
\end{figure*}

A template fit was performed on the CMS \PbPb data using distributions associated with different partons derived from \PYTHIA simulations. The primary goal was to investigate a potential shift in the rates between quark and gluon jets. Since \PYTHIA is a vacuum-like (unquenched) simulation, this fitting attempt is designed to capture the full extent of the observed modifications in the CMS \PbPb data by reducing the relative abundance of gluon jets in the measured inclusive jet sample. All quark flavors are combined into a single quark template, and a two-component quark-gluon template fit is applied to each measured CMS distribution. Figure~\ref{fig:tempFit_unfoldedCorr_pythia_cent3} illustrates the template fits for different centrality and jet \pt intervals. Centrality is defined as the fraction of the total nucleus-nucleus cross section, with the interval 0--10\% corresponding to the sample with the highest average overlap of the colliding nuclei. The fit range is shown in the \DRaxis distributions, and the lower panels display the ratio of data to fits.

\subsection{\label{sec:shifpt}Estimation of energy loss via \texorpdfstring{\pt}{pT}  shift  for inclusive, quark, and gluon jets} 
The jets reconstructed experimentally in heavy-ion collisions have been subjected to medium-induced effects (i.e., quenching), and therefore, the final-state reconstructed energy is expected to be less than that of the parton initiating the jet. Since the jet axis decorrelation observable, \DRaxis, strongly depends on the initial parton \pt, the observed narrowing of the distribution could be interpreted as a momentum shift in the reconstructed jet \pt due to jet quenching. We investigate this hypothesis by systematically increasing the jet \pt in {\PYTHIA} to find the best description of the data without varying the expected quark and gluon jet abundances (the fraction of gluon-initiated jets naturally decreases with jet \pt in the kinematic range of this study). The optimal shift is determined by the minimum $\chi^{2}$ between the \pt-shifted {\PYTHIA}-based inclusive jet \DRaxis distributions and the CMS \PbPb \DRaxis data across the nominal \pt ranges and centrality bins. This procedure involves incrementally increasing the jet \pt by $+1$ \GeV in the \PYTHIA templates at each step. For instance, for inclusive jets in the \PbPb data with \pt in the range $120 < \pt < 130$ \GeV, the corresponding \PYTHIA jet \pt range is shifted to $121 < \pt < 131$ \GeV, $122 < \pt < 132$ \GeV, and so on, with the $\chi^{2}$ value computed at each step.

\begin{table*}[htbp]
\caption{\label{tab:shiftedpTRange} Summary of the nominal \pt, shifted inclusive \pt, and shifted quark/gluon jet \pt ranges, along with their corresponding means, obtained from the \PYTHIA simulation. The shifted \pt ranges and means are determined by minimizing the $\chi^{2}$ between the \DRaxis distributions from \PYTHIA and CMS data in the 0--10\% centrality bin.}
\begin{ruledtabular}
\begin{tabular}{lcccc}
  \makecell{Nominal \pt \\ (in \GeV)} & $120 < \pt < 150$ & $150 < \pt < 190$ & $190 < \pt < 230$ & $230 < \pt < 300$ \\
  \noalign{\vskip 5pt} 
  \makecell{Nominal $\left<\pt\right>$ \\ (in \GeV)} & $131.98 \pm 0.016$ & $165.50 \pm 0.02$ & $206.20 \pm 0.024$ & $255.50 \pm 0.036$ \\
  \noalign{\vskip 5pt}
  \makecell{Nominal $\left<\pt\right>$ of \\ quark-like jets \\ (in \GeV)} & $132.11 \pm 0.022$ & $165.68 \pm 0.028$ & $206.39 \pm 0.031$ & $255.81 \pm 0.047$ \\
  \noalign{\vskip 5pt}
  \makecell{Nominal $\left<\pt\right>$ of \\ gluon-like jets \\ (in \GeV)} & $131.87 \pm 0.022$ & $165.32 \pm 0.029$ & $205.97 \pm 0.035$ & $255.05 \pm 0.056$ \\
  \noalign{\vskip 5pt}
  \makecell{Shifted \pt \\ (in \GeV)} & $142 < \pt < 172$ & $180 <\pt < 220$ & $222 <\pt < 262$ & $275 < \pt < 345$ \\
  \noalign{\vskip 5pt}
  \makecell{Shifted $\left<\pt\right>$ \\ (in \GeV)} & $154.26 \pm 0.017$ & $196.03 \pm 0.023$ & $238.58 \pm 0.026$ & $301.41 \pm 0.042$ \\
  \noalign{\vskip 5pt}
  \makecell{Shifted \pt of \\ quark-like jets  \\ (in \GeV)} & $123 < \pt < 153$ & $150 < \pt < 190$ & $192 < \pt < 232$ & $230 < \pt < 300$ \\
  \noalign{\vskip 5pt}
  \makecell{Shifted $\left<\pt\right>$ of \\ quark-like jets  \\ (in \GeV)} & $135.13 \pm 0.023$ & $165.68 \pm 0.028$ & $208.41 \pm 0.031$ & $255.81 \pm 0.047$ \\
  \noalign{\vskip 5pt}
  \makecell{Shifted \pt of \\ gluon-like jets \\ (in \GeV)} & $138 < \pt < 168$ & $177 < \pt < 217$ & $219 < \pt < 259$ & $277 < \pt < 347$ \\
  \noalign{\vskip 5pt}
  \makecell{Shifted $\left<\pt\right>$ of \\ gluon-like jets \\ (in \GeV)} & $150.12 \pm 0.024$ & $192.85 \pm 0.033$ & $235.45 \pm 0.039$ & $303.05 \pm 0.067$ \\
\end{tabular}
\end{ruledtabular}
\end{table*}

The results of this study demonstrate that applying a \pt shift to the momenta of inclusive jets produces a narrowing of the \DRaxis distributions. While the inclusive \pt-shift model exhibits greater tension with the data at low \DRaxis compared with the other methods, it provides a consistently better description in the range $\DRaxis \approx 0.03$–$0.05$, where the other methods show an apparent tension in the two higher jet \pt bins (see Fig.~\ref{fig:unfoldedCorr_pythia_shifted_mediumqg}). Nevertheless, because the data uncertainties increase with \DRaxis, the inclusive \pt-shift model shows substantially worse overall agreement with the data in terms of the $\chi^{2}$ per degree of freedom than that achieved with the two-component fit in central collisions. The average \pt shift values, representing the amount of energy lost by the jets, are found to increase with centrality, reaching the largest values in the most central 0--10\% collisions. The extracted shift values decrease progressively toward more peripheral collisions. In the 50--80\% centrality range, the nominal and shifted \pt ranges become nearly identical. Similar trends are observed when allowing for the independent variation of quark- and gluon-initiated jet energies, with gluon jets consistently preferring larger \pt-shift values. The average \pt shift for inclusive jets (i.e., with identical shifts applied to both the quark and gluon jets), as well as for independently shifted quark- and gluon-initiated jets, is presented in Table~\ref{tab:shiftedpTRange} for the 0--10\% central collisions, where the greatest energy loss is expected. As anticipated, gluon-initiated jets exhibit a larger \pt shift, consistent with larger energy loss for gluons traversing the medium. These average \pt shifts serve as a quantitative measure of the energy lost by jets during their interactions with the QGP medium.

\section{\label{sec:result}Results and discussion}
\begin{figure*}[htbp]
\centering
\includegraphics[width=0.99\textwidth]{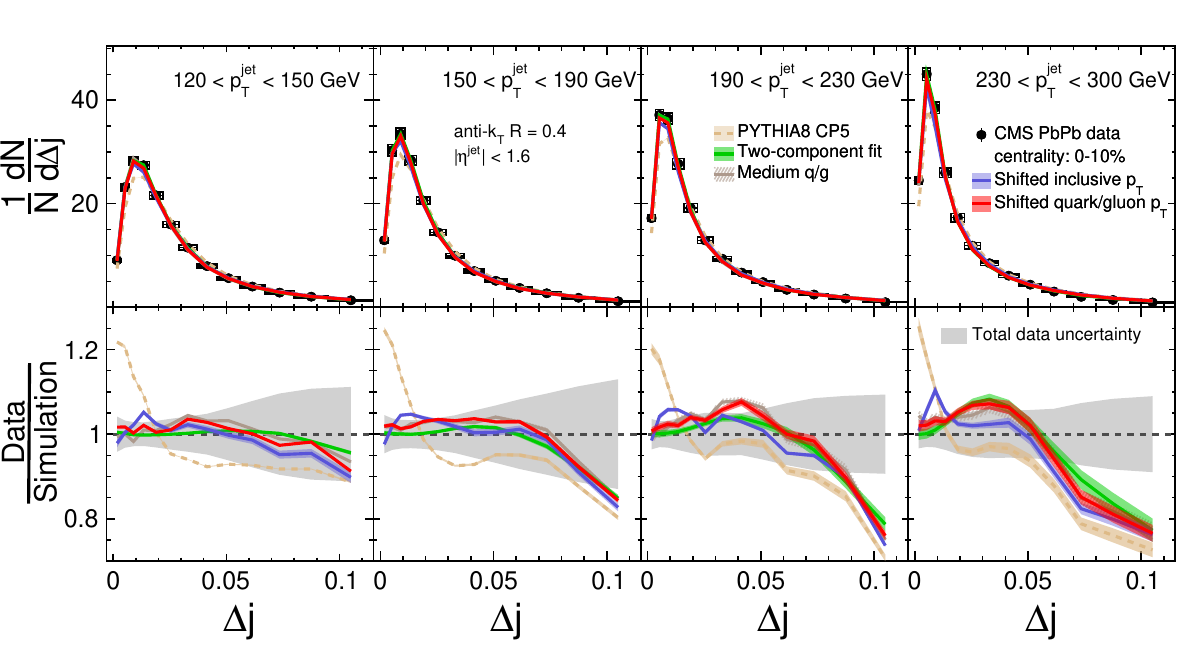}
\caption{\label{fig:unfoldedCorr_pythia_shifted_mediumqg}The normalized \DRaxis distributions from {\PYTHIA}-based simulations using shifted inclusive \pt (blue bands) and shifted quark/gluon \pt (red bands) are compared with CMS unfolded measurements in \PbPb collisions at 0--10\% centrality (black circles). The results are also compared with unquenched \PYTHIA simulations (yellow dashed-line bands), a two-component fit (green bands), and the medium q/g model (hatched lines) across various \pt intervals. Only statistical uncertainties are shown for the simulation distributions. For the CMS data points, systematic uncertainties are indicated by rectangles (correlated uncertainties) and shaded areas (uncorrelated uncertainties), while vertical solid lines depict statistical uncertainties. In the lower panels, the grey bands correspond to the total uncertainties in the data.}
\end{figure*}

Figure~\ref{fig:unfoldedCorr_pythia_shifted_mediumqg} compares the resulting template fit curves (represented by green bands), referred to as the ``two-component fit'', with the CMS \DRaxis distributions (black circles) for 0--10\% centrality across various \pt intervals. The ratios between the CMS data and simulation are displayed in the lower panels. The two-component fits capture the data trends at low \pt, but the fit quality deteriorates in higher \pt bins. It is worth noting that the discrepancies are most pronounced at high \DRaxis values, where little to no shape difference remains between the quark and gluon jet templates. These deviations suggest modifications beyond simple changes in the gluon fraction or simplistic \pt shifts are present in the data that cannot be captured by the vacuum \PYTHIA templates.

The figures also include unquenched \PYTHIA distributions (represented by the yellow dashed lines) and the modeling curve based on medium q/g fractions (shown as hatched lines). The unquenched \PYTHIA distributions exhibit the largest discrepancies with CMS data. The medium q/g modeling curve is constructed using \PYTHIA-based templates with modified quark/gluon fractions, incorporating medium-induced modifications to jet functions~\cite{CMS:2024koh, mediumqg}. This model describes the most central data points (for which predictions were available) across most of the measured \DRaxis range. However, at higher \DRaxis values in the higher \pt bins, the model overestimates the width of the data distributions, even when considering predominantly narrow (quark-initiated) jets in simulation. This observation supports the conclusion that while the bulk of the distribution is well described, the large-\DRaxis behavior observed in CMS data cannot be explained solely by changes in the quark/gluon composition of the sample. Instead, this discrepancy highlights the need to account for medium-induced substructure modifications to more accurately describe the data.

Using the shifted \pt ranges obtained through minimum $\chi^{2}$ optimization, we recalculated the \DRaxis distributions in \PYTHIA and compared them with the CMS \PbPb measurements for 0--10\% centrality across four \pt intervals (shown as the blue and red bands in Fig.~\ref{fig:unfoldedCorr_pythia_shifted_mediumqg}). Similar to the template fit curves, the \pt-shifted distributions align well with peripheral collision data within uncertainties but fail to fully reproduce the features of central collisions, particularly in the higher \pt bins at larger \DRaxis values. Interestingly, the \pt-shifted distributions closely match predictions from the two-component fit and medium q/g models, with better agreement observed for the independent shift scenario.

\subsection{\label{subsec:gfraclimit}Limits on gluon fraction}
\begin{figure*}[htbp]
\centering
\includegraphics[width=0.49\textwidth]{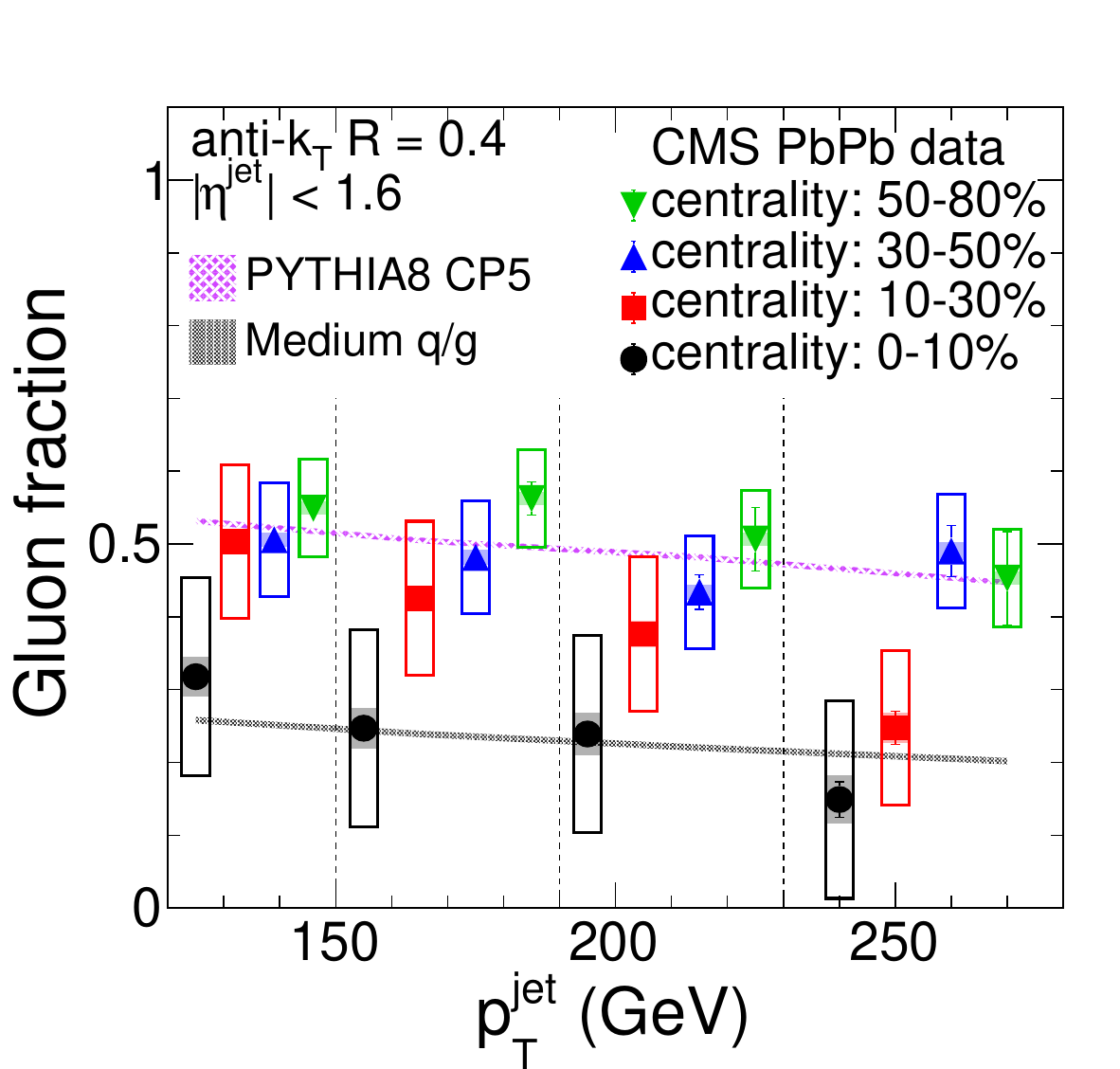}
\includegraphics[width=0.49\textwidth]{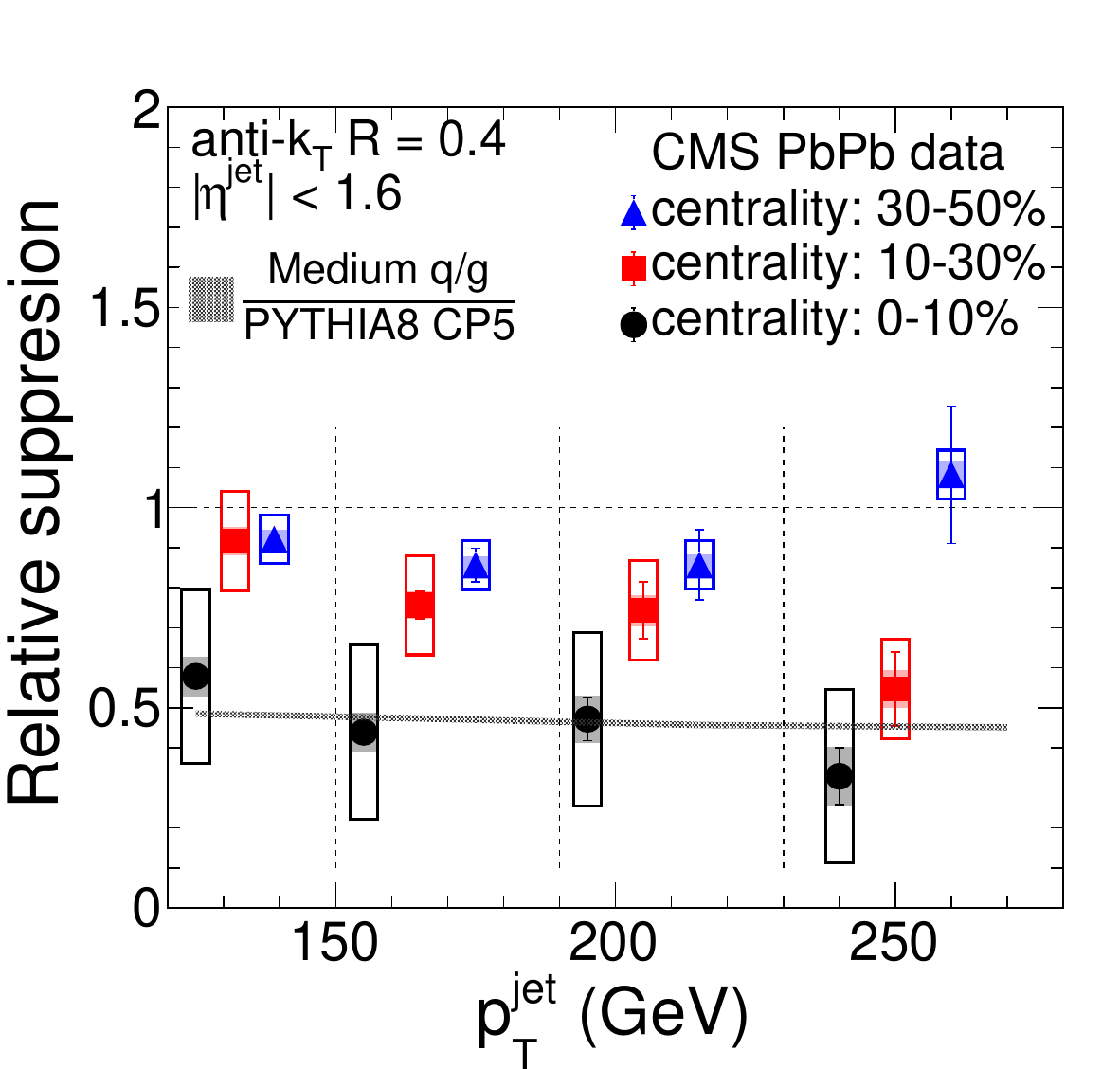}
\caption{\label{fig:gluonFrac}Left: Limits on gluon jet fraction obtained from template fits to the CMS \PbPb inclusive \DRaxis measurement using a \PYTHIA-based templates for the 0--10\% (black circles), 10--30\% (red squares), 30--50\% (blue triangles), and 50--80\% (green triangles) centrality intervals. The purple and black bands represent the gluon fraction expectations from the unquenched \PYTHIA and medium q/g models, respectively. Right: The ratio of the extracted gluon fraction limits for each centrality interval to those measured in the 50--80\% centrality interval is shown. The black band represents the ratio between the expectations from the \PYTHIA and medium q/g models. Vertical lines represent statistical uncertainties, while systematic uncertainties are shown as rectangles (correlated uncertainties) and shaded areas (uncorrelated uncertainties). Points in each jet \pt bin are shifted horizontally for better visibility.}
\end{figure*}

The left panel of Fig.~\ref{fig:gluonFrac} shows the extracted best-fit gluon fractions as a function of jet \pt, derived from the template fits to the CMS data,  presented earlier (see  Fig.~\ref{fig:tempFit_unfoldedCorr_pythia_cent3}). Since the template fitting attempts to capture all possible modifications by mere q/g fraction changes, the extracted gluon-like jet fraction values provide the lower bounds for what might be expected from realistic energy loss mechanisms. Any additional mechanisms contributing to the narrowing of the \DRaxis distribution (unrelated to changes in quark and gluon jet abundances due to color-charge effects on energy loss) would likely increase this estimate.

Each variation of the \DRaxis distributions described in Section 5 of the CMS paper~\cite{CMS:2024koh} was used to estimate the systematic uncertainties associated with the extraction of gluon-like jet fractions. For each variation, the modified \DRaxis distribution was fitted, and the corresponding gluon-jet fraction was extracted. Deviations from the nominal value were treated as systematic uncertainties. When multiple variations were available for a given source, the largest deviation from the nominal was taken. The total uncertainty was calculated by combining the individual contributions in quadrature.

The right panel of Fig.~\ref{fig:gluonFrac} presents the ratio of the gluon fraction limits for each centrality interval relative to those in the 50--80\% centrality range, emphasizing the centrality-dependent evolution of the observed variations. A lower limit on the gluon fractions is obtained for more central events, particularly at higher transverse momenta. This trend is more apparent in the ratio of the limits, where gluon-like jet fraction is reduced by approximately 50\% for 0--10\% central events. We find that the estimated gluon fraction from template fitting to the \DRaxis distributions is in agreement with the predictions from the medium q/g model for 0--10\% central collisions~\cite{CMS:2024koh, mediumqg}. 

\subsection{\label{subsec:jetraa}Calculation of jet \texorpdfstring{\Raa}{Raa} using shifted \texorpdfstring{\pt}{pt}}
\begin{figure*}[htbp]
\centering
\includegraphics[width=0.49\textwidth]{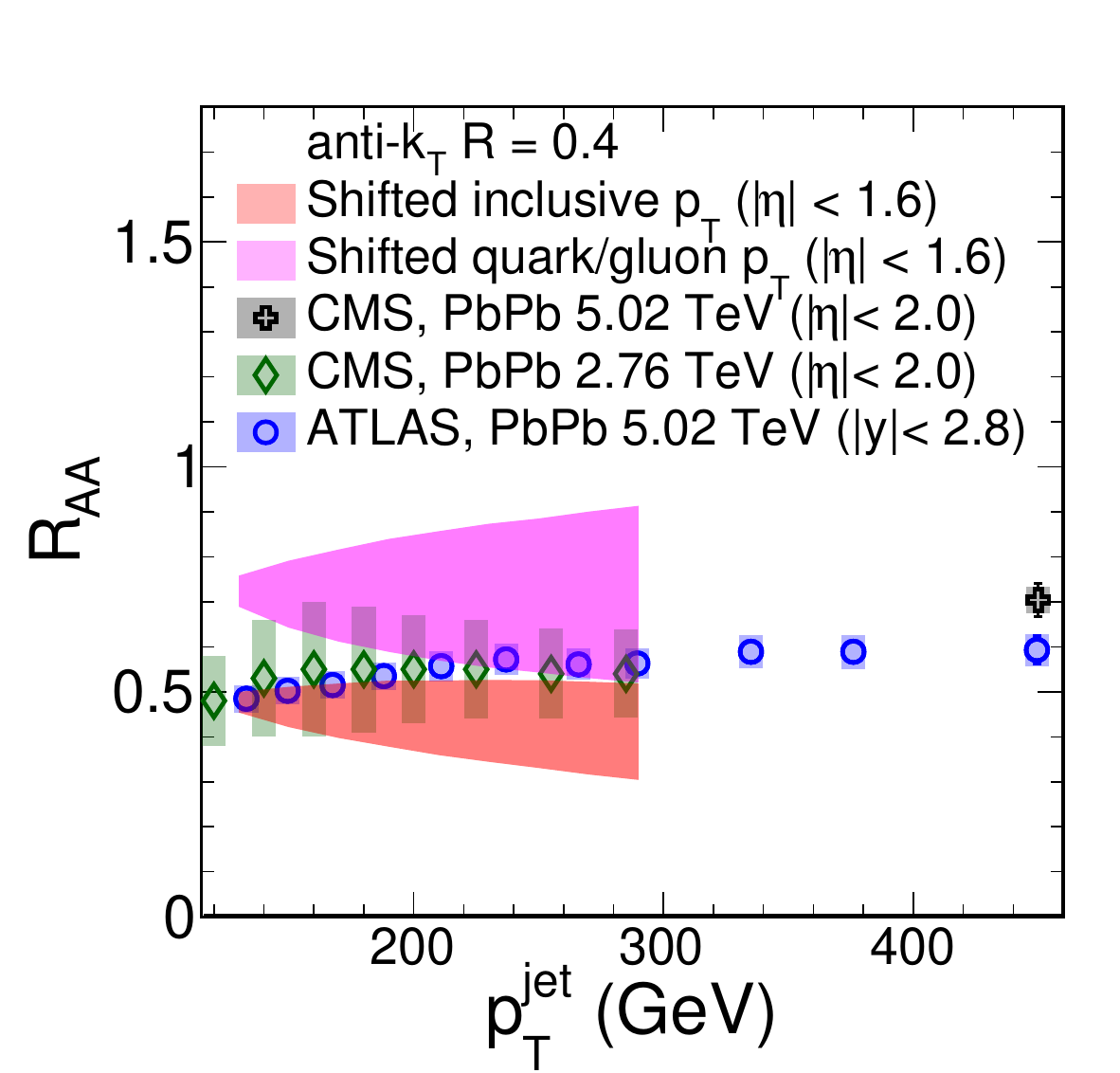}
\includegraphics[width=0.49\textwidth]{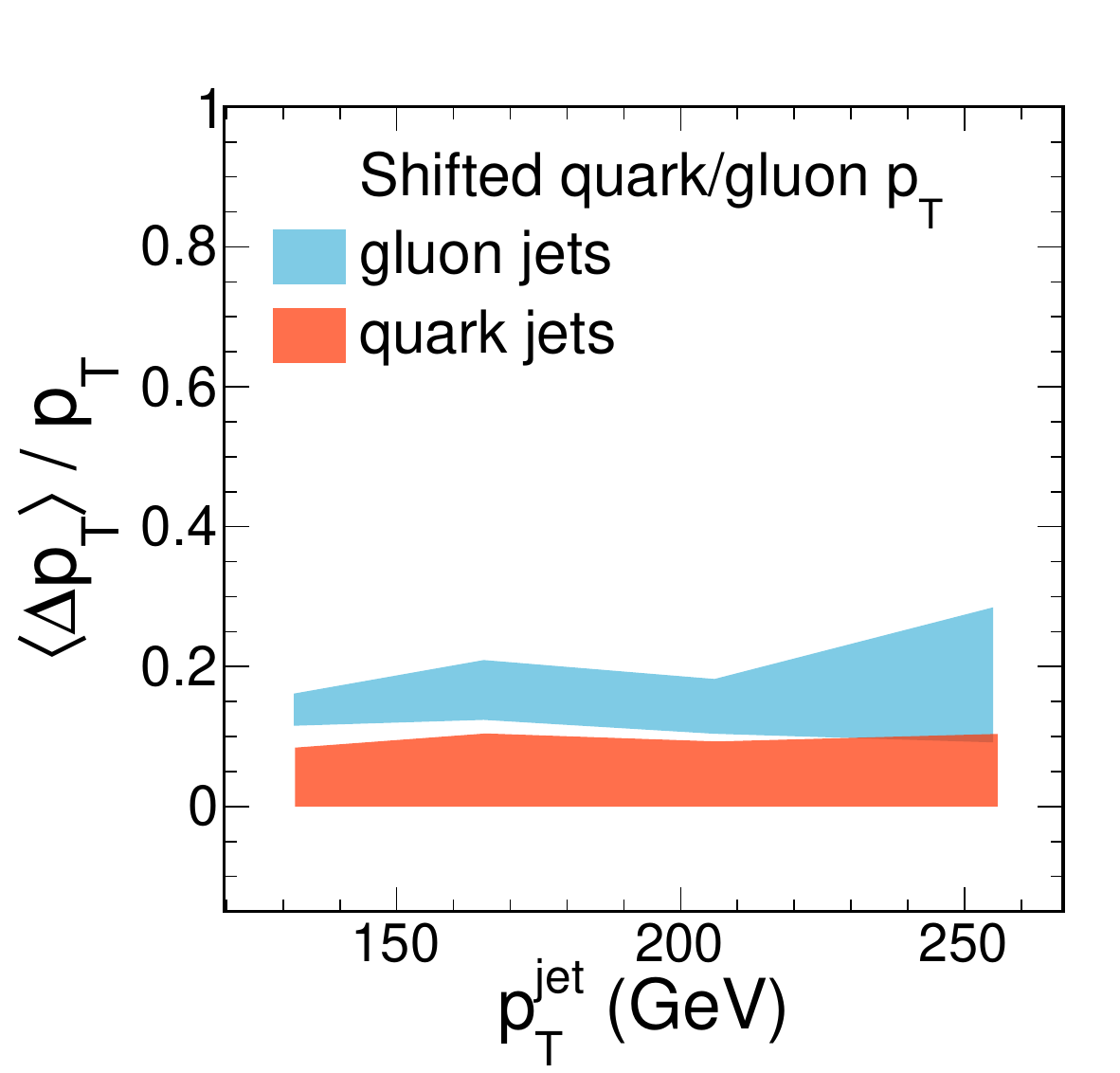}
\caption{\label{fig:RAA_dptpt}Left: The jet \Raa estimates based on inclusive jet \pt shifts along with the quark and gluon jet \pt shift values extracted from template fits in this analysis, are shown alongside other measurements and predictions (see text), for 0--10\% centrality. Right: The average fractional transverse momentum loss of quark and gluon jets, determined from the \pt shift values extracted for 0--10\% \PbPb collisions.}
\end{figure*}

To further explore the applicability of the \pt shift estimates for jet quenching modeling, we construct a jet \Raa estimate by dividing the jet counts in the shifted \pt ranges by those in the corresponding nominal \pt ranges in \PYTHIA. For the quark and gluon jet \pt shifts, jets were first counted separately for each jet type, then combined, and the resulting counts were divided by those from the nominal \pt ranges. The jet \Raa estimate constructed using the inclusive jet \pt shift is expected to provide a lower limit, while modeling the individual shifts to quark and gluon jet energies offers a more realistic estimate of energy loss. The resulting \Raa estimates, shown in the left panel of Fig.\ref{fig:RAA_dptpt}, are presented for both the inclusive jet \pt shift and the independent quark/gluon jet \pt shifts. The uncertainty on the \Raa estimate is quantified by varying the  \pt shifts by minimizing $\chi^{2}+1$ and then calculating the difference between the resulting values. Remarkably, the constructed jet \Raa estimates for both types of energy loss modeling align closely with previously measured jet \Raa from the CMS and ATLAS experiments\cite{CMS:2016uxf,CMS:2021vui,ATLAS:2018gwx}. The agreement of the estimated quenching strength determined in this study, based on the jet axis decorrelation distributions, with earlier jet \Raa measurements provides support for the plausibility of the approach developed. The proposed studies of jet-axis-based substructure observables can thus offer additional constraints on medium-induced quark/gluon jet fraction modifications, leading to a more accurate assessment of energy loss in the medium.

\subsection{\label{subsec:dptpt}Average fractional transverse momentum loss \texorpdfstring{$(\left<\Delta\pt\right>/\pt)$}{dpt/pt}}
Alongside the \Raa estimates, we also present the average fractional transverse momentum loss, $\langle \Delta\pt \rangle / \pt$, separately for quark and gluon jets, based on the \pt shift values extracted for central 0--10\% \PbPb collisions, as summarized in Table~\ref{tab:shiftedpTRange}. The uncertainties on these values were estimated in the same manner as described earlier: by minimizing $\chi^2 + 1$ and calculating the difference between the resulting values. The calculated $\langle \Delta\pt \rangle / \pt$ as a function of \pt is shown in the right panel of Fig.~\ref{fig:RAA_dptpt}. The results closely align with those reported in Ref.~\cite{eg:partonloss}, which employed a Bayesian analysis of multiple measurements across the studied \pt ranges. 

\section{\label{sec:sum}Summary}
This paper presents a phenomenological study of jet axis decorrelation (\DRaxis) between two types of jet axes for inclusive jets from lead-lead (\PbPb) collisions at a center-of-mass energy of 5.02 \TeV previously measured by the CMS experiment for different centrality selections and jet transverse momentum (\pt) intervals. The fits to these distributions are performed using templates generated with the {\PYTHIA}8 event generator for gluon-initiated jets and jets from quarks of different flavors, attempting to understand the nature of the \DRaxis modifications observed. 

The study is based on anti-\kt jets with a radius parameter $R = 0.4$  across four transverse momentum (\pt) intervals ranging from 120 to 300 \GeV within pseudorapidity $|\eta| < 1.6$. 

It is found that the (unquenched) {\PYTHIA}8 predictions describe the peripheral CMS \PbPb data but fail to capture the distributions from central events, due to the progressive narrowing of the \DRaxis distributions observed toward more central collisions. Limits on possible variations in the jet sample composition were determined using {\PYTHIA}-based two-component template fits to the CMS PbPb data. The fits result in a reduced fraction of gluon jets in the inclusive jet sample studied by CMS, with the most significant reduction observed for the highest-\pt jets in the most central events. This observation is consistent with the expected higher in-medium energy loss of gluon-initiated jets compared to those originating from hard-scattered quarks.
	
Furthermore, two possible scenarios of jet energy loss were investigated using transverse momentum shifted selections of jets in MC samples. First, an inclusive \pt shift was applied to all jets (regardless of flavor), and then differences in energy loss between quarks and gluons were probed by independently shifting the momenta of quark- and gluon-initiated jets in {\PYTHIA}8 simulations. The study was performed for the most central CMS data, where medium-induced modification to the jet-axis decorrelation distributions is most significant. The results of the inclusive \pt-shift study do not provide the best description of the measured distributions, indicating a disfavoring of a common energy loss scenario. The results of the independent quark- and gluon-jet energy variations study indicate a higher energy loss for gluon jets compared to quark jets, consistent with theoretical expectations. The extracted \pt shifts corresponding to the best description of the jet substructure data provide estimates of flavor-dependent energy loss as a function of jet \pt. The gluon jet fractions extracted from this approach are found to be reduced by amounts similar to those from the vacuum expectations as estimated from the two-component fit. To further assess the plausibility of our approach and assumptions, these \pt shifts derived from CMS central \PbPb collision data, the corresponding estimate for a jet nuclear modification factor \Raa was calculated and shown to be consistent with the measured jet \Raa, providing a coherent jet quenching picture. Finally, the average fractional transverse momentum loss for jets initiated by different partons was calculated and compared with previous experimental and theoretical results.

The study presented in this work illustrates the use of the jet substructure observable to study jet quenching mechanisms in greater detail than afforded by the inclusive-type measurements. The extracted limits on changes in the gluon-initiated jet fraction, along with estimates of quark and gluon jet energy loss in the studied kinematic range, provide new constraints on the color-charge dependence of energy loss and offer valuable insights for jet quenching models.

\begin{acknowledgments}
The authors acknowledge Austin Baty, Hannah Bossi, Manuel Calderon De La Barca, Petar Maksimovic, Sevil Salur, and Leticia Cunqueiro Mendez for useful discussions and input during the development of this work.  The authors thank Felix Ringer and Nobo Sato for providing the data points for Ref.~\cite{mediumqg}. This work is supported by the U.S. Department of Energy, grant No. DE-FG02-94ER40865.
\end{acknowledgments}

%\clearpage
\bibliography{apsbib}

\end{document}